\documentclass[11pt]{article}
\usepackage{moriond,epsfig}

\def\PRD{{\em Phys. Rev.} D}

\def\be{\begin{equation}}
\def\ee{\end{equation}}
\def\bea{\begin{eqnarray}}
\def\eea{\end{eqnarray}}

\newcommand{\B}{{\rm B}}
\newcommand{\HH}{{\mathcal H}}
\newcommand{\Om}{\Omega}
\newcommand{\ie}{\mbox{\it i.e.~}}
\newcommand{\bk}{{\mathbf k}}
\newcommand{\bq}{{\mathbf q}}
\newcommand{\vev}[1]{\mbox{$\langle #1 \rangle $}}
\newcommand{\de}{\delta}
\newcommand{\mO}{\mathcal{O}} 
\newcommand{\De}{\Delta}

\begin{document}
\vspace*{0.5cm}
\title{IMPACT OF A CAUSAL PRIMORDIAL MAGNETIC FIELD ON THE SACHS WOLFE EFFECT }

\author{ C. BONVIN }

\address{CEA, IPhT \& CNRS , URA 2306, F-91191 Gif-sur-Yvette, France.\\
	E-mail: camille.bonvin@cea.fr}

\maketitle\abstracts{We present an analytical derivation of the Sachs Wolfe effect sourced by a primordial magnetic field, generated by a causal process, such as a first order phase transition in the early universe. As for the topological defects case, we apply the general relativistic junction conditions to match the perturbation variables before and after the phase transition, in such a way that the total energy momentum tensor is conserved across the transition. We find that the relevant contribution to the magnetic Sachs Wolfe effect comes from the metric perturbations at next-to-leading order in the large scale limit. The leading order term is strongly suppressed due to the presence of free-streaming neutrinos. We derive the neutrino compensation effect and confirm that the magnetic Sachs Wolfe spectrum from a causal magnetic field behaves as $\ell(\ell+1)\,C^\B_\ell \propto \ell^2$ as found in the latest numerical analyses.}

\section{Introduction}

The origin of the large scale magnetic fields observed in galaxies and clusters is still unknown: one of the possible explanations is that they have been generated in the primordial universe. A primordial magnetic field of the order of the nanoGauss could leave a detectable imprint in the cosmic microwave background (CMB) anisotropies~\cite{Kahniashvili:2006hy,Yamazaki:2008gr,PFP,Shaw:2009nf}. Here we concentrate on its effect on the temperature CMB spectrum at large scales and more particularly on the Sachs Wolfe effect. The motivation is that conflicting results exist in the literature regarding the $\ell$-dependence of the Sachs Wolfe effect induced by a causal primordial magnetic field: the analytical analysis of \cite{Kahniashvili:2006hy} found $\ell(\ell+1)\,C^\B_\ell$ scaling as $ \ell^{-1}$ or more negative, and the same result was found in the numerical calculation of \cite{Yamazaki:2008gr}; on the other hand, \cite{PFP,Shaw:2009nf,Caprini:2009vk} found $\ell(\ell+1)\,C^\B_\ell$ scaling as $\ell^2$. The aim of this paper is to explain this discrepancy analytically (for a more detailed derivation see~\cite{cmbmag}).

We assume that a magnetic field is generated in the early universe by a sudden phase transition, as for example the electroweak (EW) phase transition. We consider a stochastic magnetic field with no background component and we suppose that the magnetic energy momentum tensor is first order in perturbation theory. We study the effect of the magnetic field on the metric and fluid (matter plus radiation) perturbations by solving analytically Einstein's and conservation equations in the long wavelength limit. We take into account the neutrinos in our derivation. In order to connect the solutions before and after the magnetic field generation, we match the geometry and the fluid variables at the phase transition time, so that the induced three metric and the extrinsic curvature are continuous~\cite{Deruelle:1997py}. This implies the conservation of the total energy momentum tensor across the phase transition, and it completely determines the metric and fluid perturbation variables after the magnetic field generation. Before neutrino decoupling, we find that at leading order in the large scale expansion $k/\mathcal{H}\ll 1$, the metric perturbation $\Phi$ is proportional to $\Phi \propto \Pi_\B (\HH/k)^2$, where $\Pi_\B$ is the magnetic field anisotropic stress. This induces a contribution in the CMB spectrum scaling as $\ell(\ell+1)\,C^\B_\ell\propto  \ell^{-1}$, and consistent with~\cite{Kahniashvili:2006hy,Yamazaki:2008gr}. However, once neutrinos decouple and start free-streaming, they acquire a non-zero anisotropic stress, which acts to compensate and reduce the magnetic field one~\cite{Shaw:2009nf,Kojima:2009gw}. We demonstrate that this compensation drastically reduces the leading order contribution to the CMB spectrum, and that the dominant contribution becomes the one from the next-to-leading order in the $k/\mathcal{H}\ll 1$ expansion, which induces then $\ell(\ell+1)\,C^\B_\ell \propto \ell^2$, as found in~\cite{PFP,Shaw:2009nf,Caprini:2009vk}.

\section{Solutions for the metric and fluid variables}

In this section we solve for the metric and fluid perturbations.
We consider only scalar perturbations on a spatially flat Friedmann background, and we work with gauge invariant variables. We consider a stochastic primordial magnetic field with spectral index $n\geq 2$ because of its causal generation~\cite{Durrer:2003ja}: $\vev{B_i(\bk)B_j^*(\bq)}=(2\pi)^3\de(\bk-\bq) (\de_{ij}-\hat{k}_i\hat{k}_j) \, A \, k^n~,$
where $A$ is the amplitude of the spectrum. We work under the one-fluid MHD approximation, meaning that the conductivity of the universe is high, so that we can neglect the electric field. The magnetic field is characterised by its energy density $\rho_\B$, anisotropic stress $\pi_\B$ and Lorentz force $\ell_\B$ that describes the momentum exchange between the magnetic field and the primordial fluid. These quantities satisfy the relation $\rho_\B/2=\pi_\B+3\ell_\B/(2k)$, coming from momentum conservation. 
To solve for the metric and fluid perturbations, we combine the system of Einstein's and conservation equations into a second order differential equation for the gauge invariant variable $D$, corresponding to the total (matter plus radiation) density perturbation in the velocity-orthogonal slicing. The equation is
\bea
 && \ddot{D}+(1+3c_s^2 -6w)\HH \dot D +3 \HH^2 \left[ -\frac{1}{2}-4w+\frac{3}{2}w^2 +3c_s^2 +\frac{c_s^2}{3} \left( \frac{k}{\HH}\right)^2 \right] D = \nonumber \\
& & \frac{\HH^2}{1+a} \left[ 2-3w+3c_s^2-\frac{a}{1+a} -\frac{1}{3} \left( \frac{k}{\HH}\right)^2 \right] \Om_\B +
\frac{2 \HH^2}{1+a} \left[ 1+6w-3c_s^2+\frac{a}{1+a}+\frac{1}{3} \left( \frac{k}{\HH}\right)^2 \right] \Pi_\B\nonumber\\
& &+ 2 \HH ^2 \left[ -2w+3c_s^2+3w^2+\frac{w}{3}  \left( \frac{k}{\HH}\right)^2 \right] \pi_{\rm F} -2\HH w \dot\pi_{\rm F}~,\label{Dddot}
\eea
where a dot denotes derivative with respect to conformal time $\eta$, $\pi_{\rm F}$ is the anisotropic stress of the matter plus radiation fluid, $\Om_\B=\rho_\B/\rho_{\rm rad}$ and $\Pi_\B=\pi_\B/\rho_{\rm rad}$. To determine the Sachs Wolfe effect, we need to solve this equation for scales which are over the horizon at recombination. Therefore, we can drop the terms proportional to $(k/\HH)^2$.

We split the universe's evolution in three stages:  before magnetic field generation, between magnetic field generation and neutrino decoupling, and after neutrino decoupling. Before magnetic field generation, the universe is filled with matter and radiation only, for which we consider adiabatic initial conditions. Since at early time neutrinos are coupled to photons, their anisotropic stress is zero. Consequently, the two last lines of Eq.~(\ref{Dddot}) vanish and we solve 
the homogeneous equation for $D$~\cite{libroruth}. At time $\eta_\B$, the magnetic field is generated by a causal process that we assume to act `fast', \ie within one Hubble time, as for example a sudden phase transition. 
After $\eta_\B$, the second line in Eq.~(\ref{Dddot}) acts as a source for $D$. We solve it using the Wronskian method. In order to connect the solutions before and after the field generation, we match the variables in such a way that the total energy momentum tensor is conserved across the transition. Following what has been done in \cite{Deruelle:1997py} for the analogous case of the topological defects, we find that this matching imposes the continuity of the variables $\phi$ and $V$ at the phase transition (see Eqs.~(36) of~\cite{Deruelle:1997py}).
Finally, at time $\eta_\nu$, neutrinos decouple and start to free stream. They acquire a non-zero anisotropic stress $\pi_\nu$ that acts as a new source term in Eq.~(\ref{Dddot}): $\pi_{\rm F}=R_\nu\pi_\nu$ (where $R_\nu\equiv \rho_\nu/\rho_{\rm rad}=0.4$). Solving Eq.~(\ref{Dddot}) using the Wronskian method requires to know the time evolution of the neutrino anisotropic stress, $\pi_\nu$. The evolution of $\pi_\nu$ in the presence of an external constant anisotropic stress has been studied in~\cite{Kojima:2009gw} (see also~\cite{Shaw:2009nf}).
 The neutrino anisotropic stress quickly adjusts to the external one and compensates it (see appendix B of~\cite{cmbmag} for a detailed derivation of $\pi_\nu$). The final time dependence of $\pi_\nu$ is rather complicated (Eq.~(B.12) of~\cite{cmbmag}) , but it can be approximated by $\pi_\nu(y)=\frac{3\Pi_\B}{R_\nu}\left(\frac{y_\nu^2}{y^2}-1 \right)-\frac{40 a_1}{15+4R_\nu}y(y-y_\nu)$,
where $y=\eta/\eta_{\rm rec}$ and $a_1$ is related to the amplitude of the primordial potential.
With this approximation, we can solve analytically for $D$.

\section{Sachs Wolfe effect at leading order in the long wavelength expansion}

With our solution for $D$ valid through the whole universe's history, we compute the metric variables $\Phi$ and $\Psi$ and the velocity perturbation $V$, using Einstein's equations. We then split the fluid into its individual components and solve for the photon density perturbation $D_{g\gamma}$ and velocity perturbation $V_\gamma$, using standard adiabatic initial conditions. With this we compute the Sachs Wolfe contribution to the temperature anisotropy at large scales, given by (see \textit{e.g.}~\cite{libroruth})
\bea
&&\frac{\Delta T}{T}^\B(k, \eta_0) \simeq \frac{D_{g\,\gamma}(k, \eta_{\rm rec})}{4} +\Psi(k, \eta_{\rm rec})-\Phi(k, \eta_{\rm rec})\\
&&\simeq -\frac{12 \, \Big[4 y_{\rm rec}^2 (y_B - 2 y_\nu) +y_{\rm rec} (4 y_B +5 y_\nu ^2 -8y_\nu)+ 6 y_\nu^2\Big]}{y_{\rm rec}^4 (2 + y_{\rm rec})^3} \frac{\Pi_\B}{x_{\rm rec}^2 }\equiv f(y_{\rm rec},y_\nu,y_\B) \frac{\Pi_\B}{x_{\rm rec}^2 }~,\nonumber
\label{SW}
\eea
where $x_{\rm rec}\equiv k\eta_{\rm rec}\ll1$ for superhorizon scales at recombination.
This contribution is proportional to the magnetic field generation time (\textit{e.g.} $y_\B\simeq 10^{-12}$ for generation at the EW phase transition), and to the neutrino decoupling time, $y_\nu\simeq 10^{-6}$. It is therefore strongly suppressed: $f(y_{\rm rec},y_\nu,y_\B)\simeq 10^{-6}$. This contribution corresponds to the effect of the magnetic field anisotropic stress from its time of generation to the neutrino  decoupling time. The subsequent magnetic contribution to the Sachs Wolfe effect, arising from $y_\nu$ up to recombination time, is cancelled by the free-streaming neutrinos. With the above result we compute the CMB spectrum. Neglecting the integrated Sachs-Wolfe, and using that, for a causal magnetic field with $n\geq 2$, the power spectrum of the magnetic field anisotropic stress is constant in $k$ up to the damping scale $k_D$~\cite{Kahniashvili:2006hy,PFP,Shaw:2009nf}
\be
\vev{\Pi_\B(\bk)\Pi_\B^*(\bq)}=(2\pi)^3\de(\bk-\bq)|\Pi_\B(k)|^2 \quad \mbox{with} \quad |\Pi_\B(k)|^2=\bar{\Pi} \, \frac{\vev{B^2}^2}{\bar\rho_{\rm rad}^2} \, \frac{1}{k_D^3}~, \quad \mbox{we find~\cite{Caprini:2009vk}}
\label{PiBspec}
\ee
\bea
\hspace*{-0.8cm}\ell(\ell+1) C_\ell^\B &\simeq & f^2(y_{\rm rec}, y_\nu,y_\B) \, g^2(\eta_{\rm rec}) \, \bar\Pi \, \frac{\vev{B^2}^2}{\rho_{\rm rad}^2} \, \frac{\eta_0}{\eta_{\rm rec}} \,
\frac{1}{(\eta_{\rm rec} k_D)^3} \, \frac{2\, \ell(\ell+1)}{8\ell^3+12\ell^2-2\ell-3}~,
\label{Cl_leading}
\eea
where $g(\eta_{\rm rec})$ denotes the visibility function.
Therefore, we find that the CMB spectrum scales as $1/\ell$, as in~\cite{Kahniashvili:2006hy,Yamazaki:2008gr}, rather than as $\ell^2$ as found in  \cite{PFP,Shaw:2009nf}. The reason why refs.~\cite{PFP,Shaw:2009nf} do not find the $1/\ell$ dependence is because the initial conditions that they insert into their Boltzmann code are derived after neutrino decoupling, when the magnetic field anisotropic stress has already been compensated by the neutrino one. Eq.~(\ref{Cl_leading}) shows that the period of time between magnetic field generation and neutrino decoupling leaves an imprint on the CMB spectrum. This imprint is however much too small to be observable. Therefore, we now proceed to compute analytically the next order contribution in the $k/\HH\ll1$ expansion, that will lead to a $\ell^2$-dependence in the CMB spectrum.

\section{Next-to-leading order contribution to the Sachs Wolfe}

In order to compute the next-to-leading order contribution to the temperature anisotropy, we need to solve for the metric variables at next-to-leading order in the $k/\HH\ll1$ expansion. The easiest way to compute this order is to use the curvature perturbation $\zeta\equiv-\Phi +\frac{2}{3(1+w)} \left( \Psi -\frac{\dot{\Phi}}{\HH} \right)$. Indeed, starting from the leading order solution that we computed for $D$, we can calculate the next-to-leading order in the curvature \footnote{Note that the magnetic field does not affect the curvature at leading order, $\mO(1/x_{\rm rec}^{2})$. It is therefore still conserved on large scales.}. We use then this solution to compute the next-to-leading order in $\Phi$, by integrating from the definition of $\zeta$. This can be done analytically if we approximate the time evolution of the neutrino anisotropic stress at next-to-leading order as $\pi_\nu(y)=(d_1\Om_\B+d_2\Pi_\B)(y-y_\nu)^2 x_{\rm rec}^2
$, where $d_1$ and $d_2$ are two arbitrary constants that we determine from the conservation equations of the neutrino and photon fluids. With this, we find for the temperature anisotropy at next-to-leading order
\be
\frac{\De T}{T}^\B (k, \eta_0) \simeq -0.2 \, \Om_\B - 2.7 \, \Pi_\B~.
\ee
The energy density spectrum and the cross-correlation one have been calculated in~\cite{Kahniashvili:2006hy,PFP,Shaw:2009nf}, and they share the same $k$-dependence as the anisotropic stress spectrum, Eq.~(\ref{PiBspec}). Denoting $\bar{\Om}$ the amplitude of the energy density spectrum and $\bar{C}$ the cross-correlation one, we find for the CMB spectrum
\bea
\hspace*{-1cm}\ell(\ell+1) C_\ell^\B 
&\simeq & g^2(\eta_{\rm rec}) \, \Big[ 0.04 \, \bar\Om +7.29 \,  \bar\Pi +0.54  \, \bar{C} \Big] \, \frac{\vev{B^2}^2}{\bar\rho_{\rm rad}^2} \, \frac{\ell(\ell+1)}{\pi\,(\eta_0 k_D)^2}~.
\eea
Therefore, we confirm that the next-to-leading order contribution to the Sachs Wolfe effect scales as $\ell(\ell+1)\,C^\B_\ell \propto \ell^2$, as found numerically in~\cite{PFP,Shaw:2009nf}.

\section{Conclusion}

In this work we present an analytical computation of the Sachs Wolfe effect induced by a primordial magnetic field. We have restricted our analysis to a magnetic field generated by a causal process, such as a first order phase transition. In this case, the initial conditions for the metric and fluid variables are determined unambiguously by imposing conservation of the total energy momentum tensor across the transition. Using these initial conditions, we have computed analytically the leading order and next-to-leading order magnetic contribution to the Sachs Wolfe effect. We have found that the leading order contribution is sourced only by the magnetic field anisotropic stress, and leads to a CMB spectrum scaling as $1/\ell$. However, this contribution is strongly suppressed once the magnetic field anisotropic stress is compensated by the one of the neutrinos. As a consequence, the dominant contribution to the Sachs Wolfe is the next-to-leading order one, that generates a CMB spectrum scaling as $\ell^2$. Our analytical work solves therefore the discrepancy regarding the $\ell$-dependence of the magnetic Sachs Wolfe.

\end{document}